\documentclass[12pt,psamsfonts]{amsart}
\usepackage{amsmath}
\usepackage{amsthm}
\usepackage{amssymb}
\usepackage{amscd}
\usepackage{amsfonts}
\usepackage{amsbsy}
\usepackage{blindtext}
\usepackage{booktabs}
\usepackage{color}
\usepackage{calc}
\usepackage{graphicx}
\usepackage{subfigure}
\usepackage{setspace}
\usepackage{tikz}
\usetikzlibrary{arrows, patterns}

\begin{document}

\title[]
{Global dynamics of the Ho\v{r}ava-Lifshitz cosmology in the presence of non-zero cosmological constant in a flat space}
 \author{Fabao~Gao$^{1,2}$, and Jaume~Llibre$^{2}$}
 \address{${}^1$School of Mathematical Science, Yangzhou University, Yangzhou 225002, China}
  \address{\textnormal{E-mail: gaofabao@sina.com (Fabao Gao, ORCID 0000-0003-2933-1017)}}
 \address{${}^2$ Departament de Matem$\grave{a}$tiques, Universitat Aut$\grave{o}$noma de Barcelona, Bellaterra 08193, Barcelona, Catalonia, Spain}
  \address{\textnormal{E-mail: jllibre@mat.uab.cat (Jaume Llibre, ORCID 0000-0002-9511-5999)}}
  

\begin{abstract}
Using the qualitative theory of differential equations, the global dynamics of a cosmological model based on Ho\v{r}ava-Lifshitz gravity is studied in the space with zero curvature in the presence of the non-zero cosmological constant.
\end{abstract}

\maketitle

\section{Introduction}
A decade ago Ho\v{r}ava \cite{Horava} brought forward a new theory on space-time asymmetric gravitation, called Ho\v{r}ava-Lifshitz gravity, together with the scalar field theory of Lifshitz. This theory's applications in cosmology, dark energy, and black hole have stimulated many studies (See review papers \cite{Mukohyama}, \cite{Sotiriou} or regular literature \cite{Abreu}-\cite{Bhattacharjee}). 

Based on whether the value of $\Lambda$ (cosmological constant) is zero and the flatness of the universe, i.e. whether the space curvature $k$ is equal to zero, Leon et al. \cite{Leon2012,Leon2019,Leon2009,Paliathanasis} divided Ho\v{r}ava-Lifshitz cosmology into four cases: (1) $\Lambda = 0, k = 0$; (2) $\Lambda = 0, k\neq 0$; (3) $\Lambda\neq 0, k = 0$; (4) $\Lambda\neq 0, k\neq 0$ under the classic FLRW metric. 
They either studied partially the three-dimensional dynamics of Ho\v{r}ava-Lifshitz cosmology, or analyzed its two-dimensional dynamics under the usual exponential potentials. 

For the cosmological constant $\Lambda$ that scientists have been concerning about, Carlip \cite{Carlip} believed that the vacuum fluctuations under the standard effective field theory produce a huge $\Lambda$ and produce high $k$ on the Planck scale, but it is almost invisible at the observable scale. Compared with the prediction in the standard $\Lambda$ cold dark matter model, Valentino et al. \cite{Valentino} proposed that the cosmological space may be a closed three-dimensional sphere, i.e., the cosmological space's curvature may be positive, based on the enhanced lensing amplitude in the cosmic microwave background power spectra confirmed by Planck Legacy 2018 release. Although this study provides the latest results, the debate about the universe's shape has not yet been settled. An important reason is that the calculation of the critical density of the universe depends on the measurement of the Hubble constant, but it seems that this constant is still not accurate, then the boundary line of the universe is fuzzy, so it is too early to say that the universe must be closed.

The global dynamics of the Ho\v{r}ava-Lifshitz cosmology under the background of FLRW with $k=0,\Lambda=0$ was studied in \cite{Gao20191}, and the case of $k\neq0,\Lambda=0$ has also been addressed in \cite{Gao20192}. In this paper we will consider the flat universe with $\Lambda\neq0$, the corresponding autonomous cosmological equations admit the following form
\begin{equation}
\begin{array}{rl} \vspace{2mm}
\dfrac{dx}{dt}&=\sqrt{6}s\left(z^2-x^2+1\right)+3x\left(x^2-1\right),\\ \vspace{2mm}
\dfrac{dz}{dt}&=3zx^2,\\ \vspace{2mm}
\dfrac{ds}{dt}&=\dfrac{\sqrt{6}}{n}xs^2.
\end{array}
\label{eq1}
\end{equation}
The details for obtaining the cosmological equations (\ref{eq1}) will be provided in Section 2.

\section{The cosmological equations}
In order to describe the cosmological model, we first briefly review the Ho\v{r}ava-Lifshitz theory of gravity proposed in \cite{Horava}. The field content in this theory can be derived from the space vector $N_i$ and scalar $N$, see \cite{Leon2009,Kiritsis}. They are actually common `lapse' and `shift' variables in general relativity. From this the complete metric can be expressed as
\begin{equation}
\begin{array}{rl} 
ds^2 = -N^2dt^2 + g_{ij}(dx^i + N^idt)(dx^j + N^jdt),\ \ \ N_i=g_{ij}N^j,
\end{array}
\end{equation}
where $g_{ij}$ is a spatial metric, here $i$ and $j$ are natural numbers from 1 to 3. The coordinate transformations follow $t\to l^3t,\ x^i\to lx^i$. Note that $g_{ij}$ is invariant, the same as $N$, but $N^i$ is scaled to $l^{-2}N_i$.\\
\indent According to the detailed-balance condition, the full gravitational action of Ho\v{r}ava-Lifshitz is expressed as
\begin{equation}
\begin{array}{rl} \vspace{2mm}
 S_g=&\displaystyle\int dtd^3x\sqrt{g}N\left\{\dfrac{2}{\kappa^2}\left(K_{ij}K^{ij}-\lambda K^2\right)-\dfrac{\kappa^2}{2w^4}C_{ij}C^{ij}\right.\\\vspace{2mm}
& \ \ \ +\dfrac{\mu\kappa^2}{2w^2}\dfrac{\epsilon^{ijm}}{\sqrt{g}}R_{il}\nabla_jR^l_k-\dfrac{\mu^2\kappa^2}{8}R_{ij}R^{ij}\\
& \ \ \ -\left.\dfrac{\mu^2\kappa^2}{8(3\lambda-1)}\left(\dfrac{1-4\lambda}{4}R^2+\Lambda R-3\Lambda^2\right)\right\},
\end{array}
\end{equation}
where $C^{ij}=\epsilon^{ijm}\nabla_k\left(4R^j_i-R\delta^j_i\right)/(4\sqrt{g})$ denotes the Cotton tensor, $K_{ij}=(\dot{g}_{ij}-\nabla_iN_j-\nabla_jN_i)/(2N)$ represents the extrinsic curvature, and $\epsilon^{ijm}/\sqrt{g}$ is the standard general covariant antisymmetric tensor, the indices are to up and down with the metric $g_{ij}$. $\kappa$, $\lambda$, $w$ and $\mu$ are all constants, for more details see \cite{Horava}.\\

\indent Consider the gravitational action term on the potential $V(\phi)$ as follows
\begin{equation}
\begin{array}{rl} \vspace{2mm}
 S=\displaystyle\int dtd^3x\sqrt{g}N\left(\dfrac{3\lambda-1}{4}\dfrac{\dot{\phi}^2}{N^2}-V(\phi)\right),
\end{array}
\end{equation}
and the metric $N^i=0$, $g_{ij}=a^2(t)\gamma_{ij}$, $\gamma_{ij}dx^idx^j=r^2d\Omega^2_2+dr^2/(1-kr^2)$. 
Above $a(t)$ represents the scale factor of the expanding universe, which is dimensionless, and $\gamma_{ij}$ refers to the constant curvature metric with maximally symmetric. For the flat space we take $k=0$ in this paper.\\
\indent 
For the sake of simplicity, $\kappa^2$ and $N$ are normalized, and then the corresponding cosmological model can be interpreted as
\begin{equation}
\begin{array}{rl} \vspace{2mm}
&H^2=\dfrac{\dot{\phi}^2}{24}+\dfrac{V(\phi)}{6(3\lambda-1)}-\dfrac{\mu^2\Lambda^2}{16(3\lambda-1)^2},\\ \vspace{2mm}
&\dot{H}+\dfrac{3}{2}H^2=-\dfrac{\dot{\phi}^2}{16}+\dfrac{V(\phi)}{4(3\lambda-1)}-\dfrac{3\mu^2\Lambda^2}{32(3\lambda-1)^2},\\ \vspace{2mm}
&\ddot{\phi}+3H\dot{\phi}+\dfrac{2V'(\phi)}{3\lambda-1}=0,
\end{array}
\end{equation}
where $H$ is the Hubble parameter and has the form $\dot{a}(t)/a(t)$.\\
\indent Considering that $V(\phi)$ admits various mathematical forms (see \cite{Leon2019,Fadragas2014,Escobar2014,Alho}), we just take $V(\phi)=(\mu\phi)^{2n}/2n$ with a natural number $n$ and a constant $\mu>0$ in this paper. Following \cite{Leon2019,Leon2009} we do the dimensionless transformation
\begin{equation}
\begin{array}{rl} \vspace{2mm}
&x=\dfrac{\dot{\phi}}{2\sqrt{6}H},
\ \ \ y=\dfrac{\sqrt{V(\phi)}}{\sqrt{6}H\sqrt{3\lambda-1}},
\ \ \ z=\dfrac{\Lambda\mu}{4(3\lambda-1)H},\\ \vspace{2mm}
&s=-\dfrac{V'(\phi)}{V(\phi)},\ \ \ f(s)\equiv\dfrac{V''(\phi)}{V(\phi)}-\dfrac{V'(\phi)^2}{V(\phi)^2}.
\end{array}
\label{eq5}
\end{equation}
Thus we obtain $f(s)=-s^2/(2n)$, which is a power-law potential, so $ds/dt=\sqrt{6}xs^2/n$. Furthermore it can be followed from equations (5) and (6) that
\begin{equation}
\begin{array}{rl} \vspace{2mm}
x^2+y^2-z^2=1,\\ \vspace{2mm}
\dfrac{H'}{H}=-3x^2.
\end{array}
\end{equation}

Therefore the field equations become the dimensionless form provided in \cite{Horava}. For more details on system (1) see the equations (205)-(207) of \cite{Leon2019} or equations (52)-(54) of \cite{Paliathanasis}. 

The present paper gives a fully description of the global dynamics of system (1) in the physical area of interest $G=\left\{(x,z,s)\in\mathbb R^3: x^2-z^2\leq1\right\}$. In sections 3 and 4 we will investigate the phase portraits of system (1) at finite and infinite equilibrium points on invariant planes and surface. In section 5 we will discuss the phase portraits of system (1) inside the Poincar\'{e} ball restricted to the region $G$. An introduction to the Poincar\'{e} ball that can be used to study the dynamics of the system (1) near infinity can be found in the appendix. 
Based on these sections, considering the symmetry of system (1), we will study the global dynamics of system (1) adding its behavior at infinity in section 6. Moreover we will give the final discussion and summary in the last section 7.

\section{Phase portraits on two invariant planes $z=0,\ s=0$ and on the invariant surface $x^2-z^2=1$}
\par In order to clarify the local phase portraits of equilibrium points (finite and infinite) and the global phase portraits of system (\ref{eq1}) in the aforementioned region $G$ (refer to \cite{Leon2019} or \cite{Paliathanasis} again). We begin to describe the phase portraits on the invariant planes $z=0,\ s=0$ and on the invariant surface $x^2-z^2=1$.
\subsection{The invariant plane $z=0$}
\par On the plane $z=0$ system (1) reduces to
\begin{equation}
\begin{array}{rl}
\dfrac{dx}{dt}&=\left(x^2-1\right)\left(3x-\sqrt{6}s\right),\\ \vspace{2mm}
\dfrac{ds}{dt}&=\dfrac{\sqrt{6}}{n}\,xs^2.
\end{array}
\label{eq9}
\end{equation}
The phase portrait of the above system in the strip $z=0$ and $x^2-z^2\leq1$ has been presented in \cite{Gao20191} (see Figure 1). System (\ref{eq9}) contains a hyperbolic equilibrium point $e_{0}=(0,0)$ and two semi-hyperbolic equilibrium points $e_{1}=(1,0)$, $e_{2}=(-1,0)$, where $e_{0}$ is a saddle point, and the other two are saddle-nodes.
\begin{figure}[htbp]
\begin{minipage}[t]{120mm}
\vspace {2mm}
\centering\includegraphics[width=5cm]{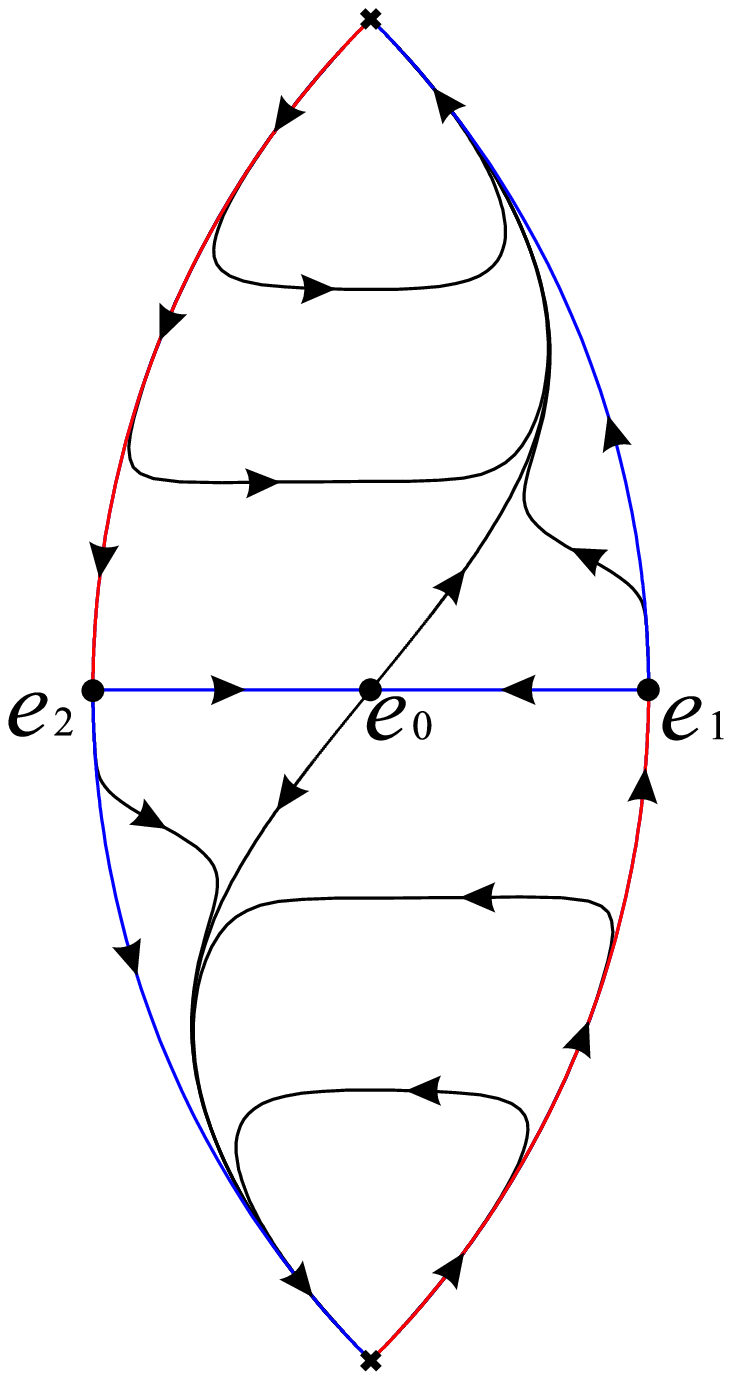}
  \subfigure{\includegraphics[width=2cm]{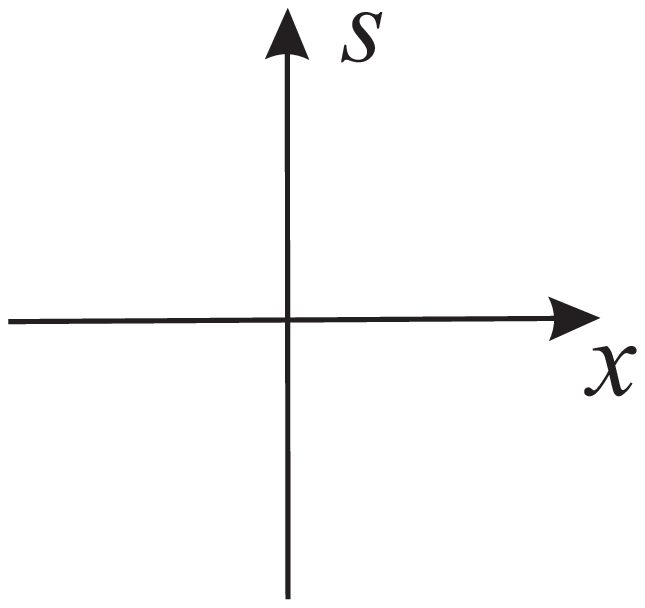}}
\caption{The phase portrait of system (\ref{eq1}) on the invariant plane $z=0$ in the region $-1\leq x\leq1$.}
\label{Fig}
\end{minipage}
\end{figure}
\subsection{The invariant plane $s=0$}
\par On the plane $s=0$ system (1) reduces
\begin{equation}
\begin{array}{rl}
\dfrac{dx}{dt}=3x\left(x^2-1\right),\ 
\dfrac{dz}{dt}=3zx^2.
\end{array}
\end{equation}
Note that the straight line $x=0$ is filled of equilibrium points. Introducing the transformation with respect to time $d\tau_1=xdt$ yields
\begin{equation}
\begin{array}{rl}
\dfrac{dx}{d\tau_1}=3\left(x^2-1\right),\ 
\dfrac{dz}{d\tau_1}=3zx,
\end{array}
\end{equation}
which has two hyperbolic equilibrium points $e_{1}=(1,0)$ and $e_{2}=(-1,0)$. Here $e_{1}$ is an unstable node and has two eigenvalues 3 and 6, but $e_{2}$ is a stable node and has eigenvalues -3 and -6.
\par According to the Poincar\'e compactification method (see Chapter 5 of \cite{Dumortier} for more details), system (9) on the local chart $U_1$ reduces to
\begin{equation}
\begin{array}{rl} 
\dfrac{du}{dt}=3uv^2,\ 
\dfrac{dv}{dt}=3v\left(v^2-1\right).
\end{array}
\end{equation}
Since all the points at infinity (i.e. at $v=0$) of system (11) are equilibrium points, we do the transformation of the time $d\tau_2=vdt$, and the system (11) becomes
\begin{equation}
\begin{array}{rl} 
\dfrac{du}{d\tau_2}=3uv,\ 
\dfrac{dv}{d\tau_2}=v^2-1.
\end{array}
\end{equation}
However there is no equilibrium points in system (12).
\par System (9) on the local chart $U_2$ becomes
\begin{equation}
\begin{array}{rl}
\dfrac{du}{dt}=-3uv^2,\ 
\dfrac{dv}{dt}=-3u^2v.
\end{array}
\end{equation}
Since this system's linear term is always equal to zero, the corresponding topological index is known to be zero by the Poincar\'e-Hopf Theorem (for more details, see Theorem 6.30 in \cite{Dumortier}). To study the local phase portrait of the equilibrium point (0,0) of system (13), we use the vertical blow-up techniques (see Ref. \cite{Alvarez}), i.e., let $w=v/u$ then we have
\begin{equation}
\begin{array}{rl}
\dfrac{du}{dt}=-3u^3w^2,\ 
\dfrac{dw}{dt}=3u^2w(w^2-1).
\end{array}
\end{equation}
Rescaling system (14)'s time $t$ by doing $d\tau_3=3u^2wdt$ yields
\begin{equation}
\begin{array}{rl}
\dfrac{du}{d\tau_3}=-uw,\ 
\dfrac{dw}{d\tau_3}=w^2-1.
\end{array}
\end{equation}
This system admits two equilibrium points $(0,-1)$ and $(0,1)$ on $u=0$. Both of these two points are hyperbolic unstable saddle points with eigenvalues of $-2,\ 1$ and $2,\ -1$, respectively. The local phase portrait around them is shown in Figure 2(a). Note the time rescaling between the above two systems, the local phase portrait of system (14) can be found in Figure 2(b). Additionally, all points on the axes $u=0$ and $w=0$ are singularities of system (14). Since $u>0$ and $w>0$ in the first quadrant I of Figure 2(b), then $v=uw$ will decrease as $u$ decreases, so the local phase portrait in the quadrant I of the $u$-$w$ coordinate system (corresponding to system (14)) can be equivalently converted to the portrait in the first quadrant of the $u$-$v$ coordinate system (corresponding to system (13)). Similarly the local phase portraits in the quadrants II, III and IV of the $u$-$w$ coordinate system can also be equivalently converted to these portraits in the third, second and fourth quadrants of the $u$-$v$ coordinate system, respectively. Therefore the local phase portrait of system (13) is displayed in Figure 2(c), and the corresponding local phase portrait at the origins of $U_2$ and the symmetrical $V_2$ in the invariant plane $s=0$ can be found in Figure 2(d).

\begin{figure}[]
  \centering
  \begin{minipage}{130mm}
  \subfigure[]{\label{fig:subfig:a}
    \includegraphics[width=6cm]{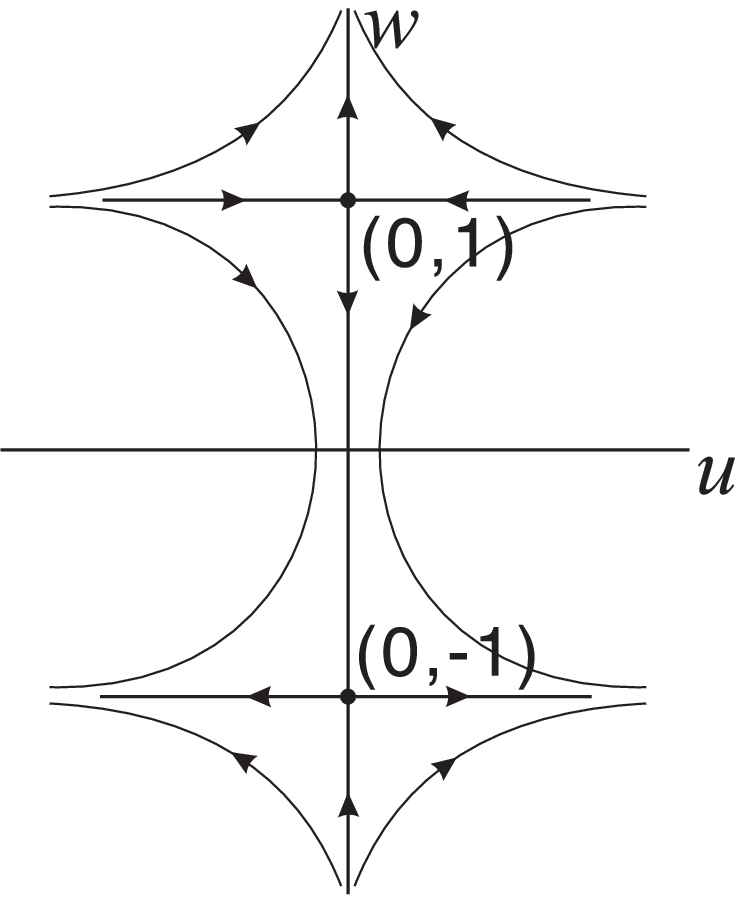}}
  \subfigure[]{\label{fig:subfig:b}
    \includegraphics[width=6cm]{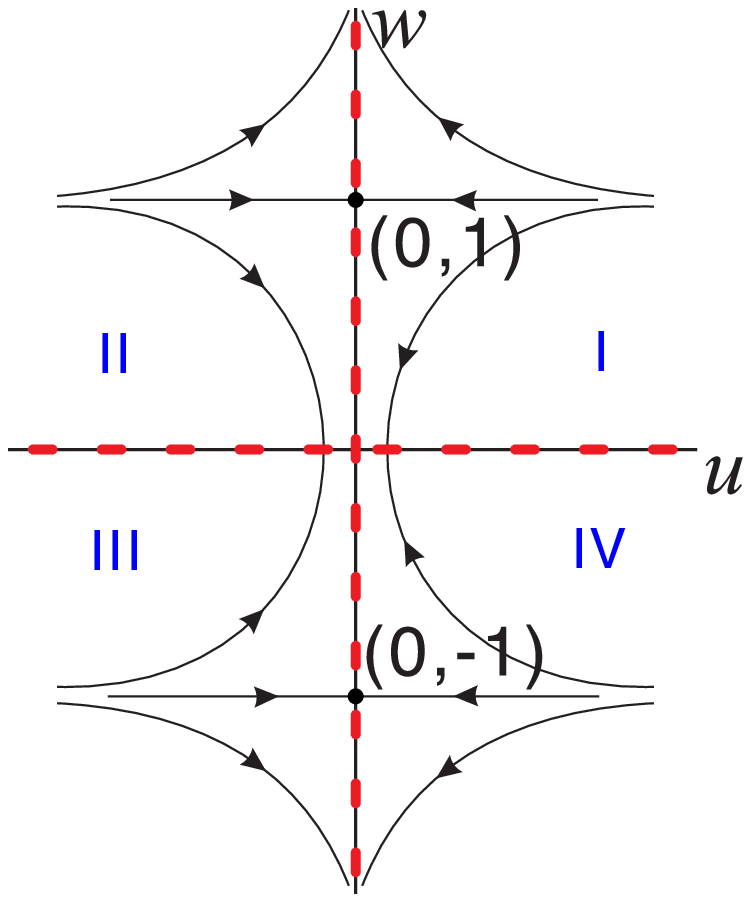}}\\
  \subfigure[]{\label{fig:subfig:c}
    \includegraphics[width=6cm]{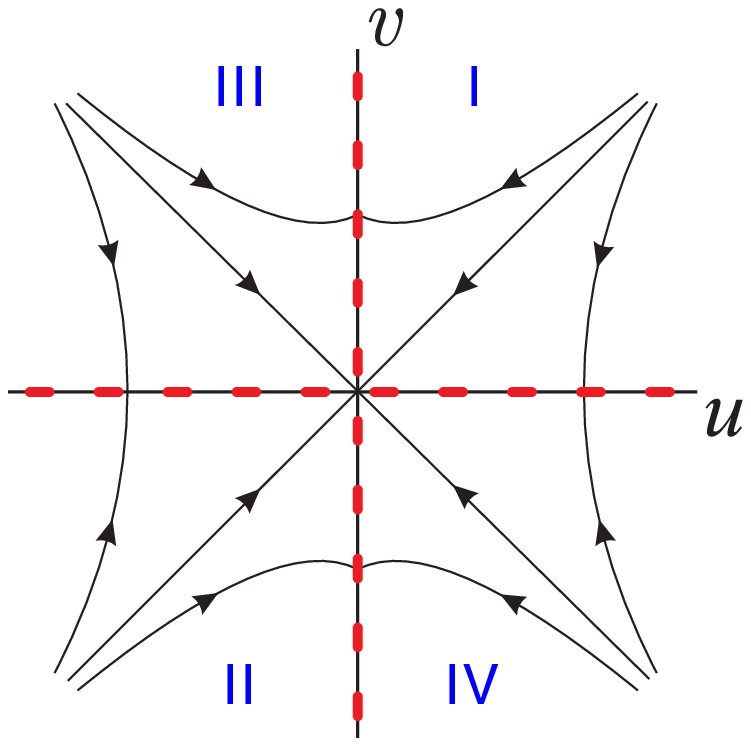}}
  \hspace{0.8mm}
  \subfigure[]{\label{fig:subfig:d}
    \includegraphics[width=5.25cm]{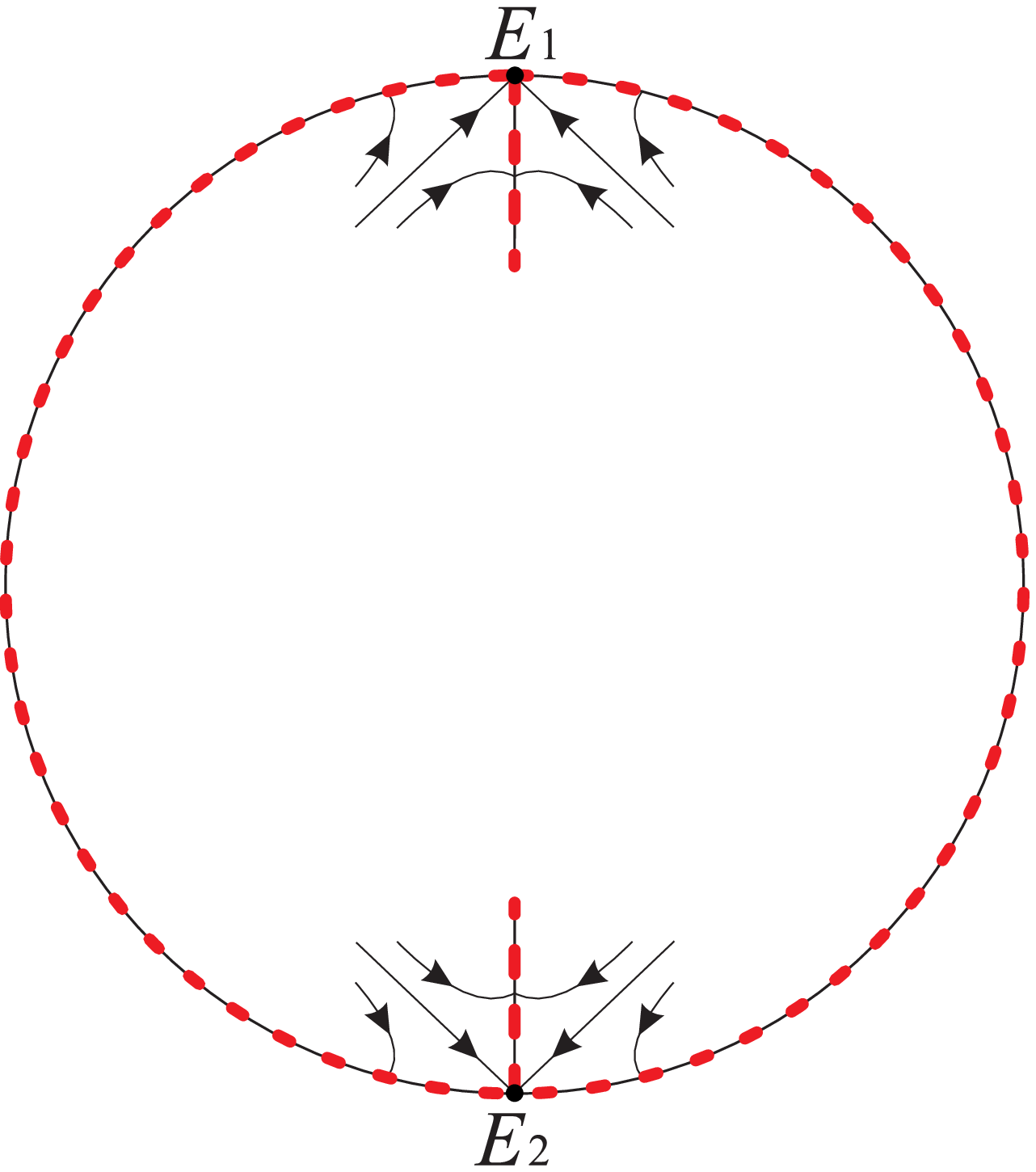}}
    \caption{In (a), (b) and (c) there are the local phase portraits of the equilibrium points of systems (15), (14) and (13), respectively. In (d) there is the local phase portraits at the origins of $U_2$ and $V_2$ for $s=0$.}
  \label{fig:subfig}
  \end{minipage}
 \end{figure}
In summary, according to the previous information and considering that the straight lines $x=0$ and $z=0$ are invariant under the flow of system (9), we can obtain that in the Poincaré disk with $s=0$, the global phase portrait is restricted to the strip $-1\leq x\leq1$ in Figure 3. 
\begin{figure}[htbp]
\begin{minipage}[t]{120mm}
\vspace {2mm}
\centering\includegraphics[width=7cm]{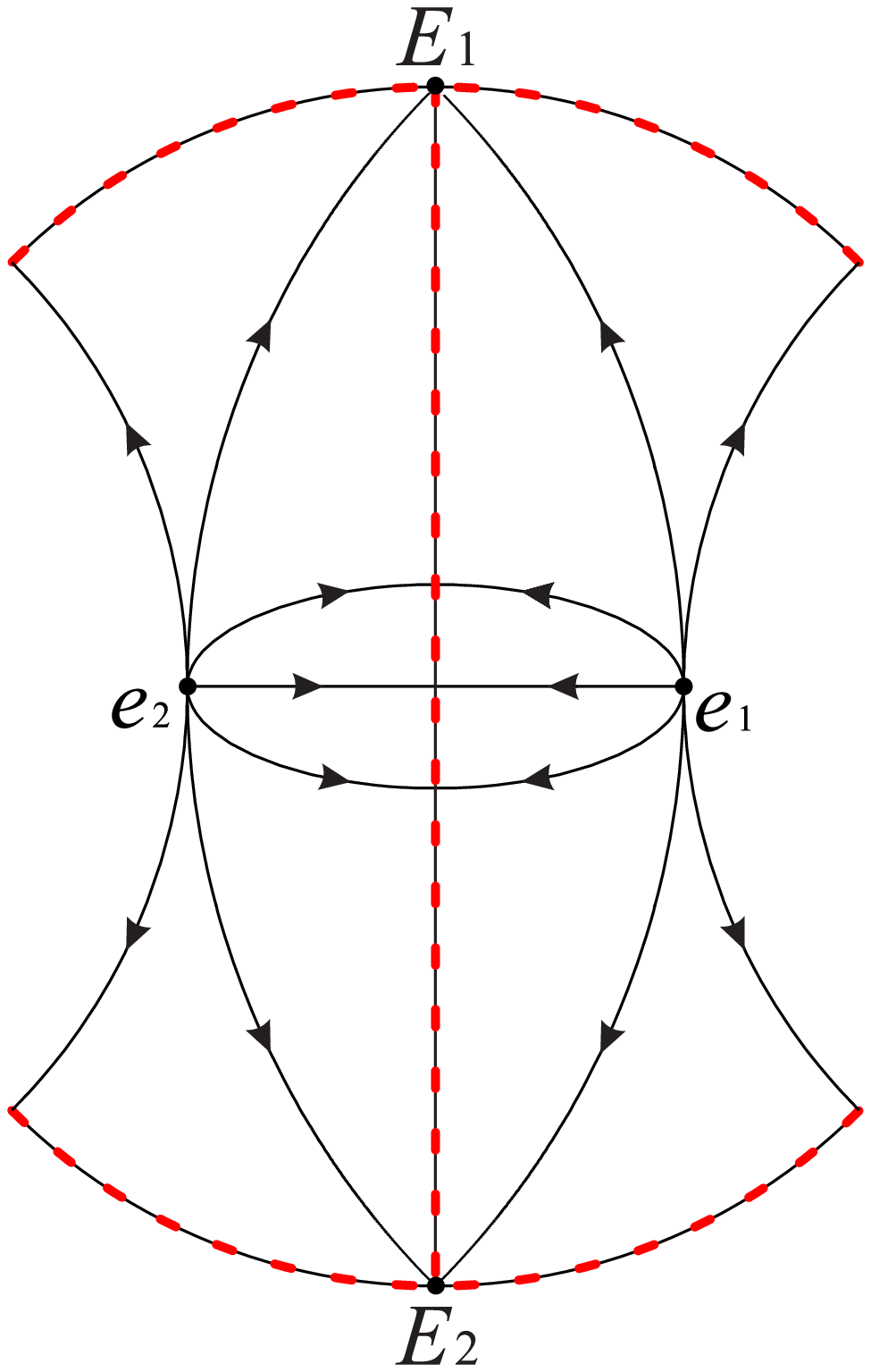}
  \subfigure{\includegraphics[width=2cm]{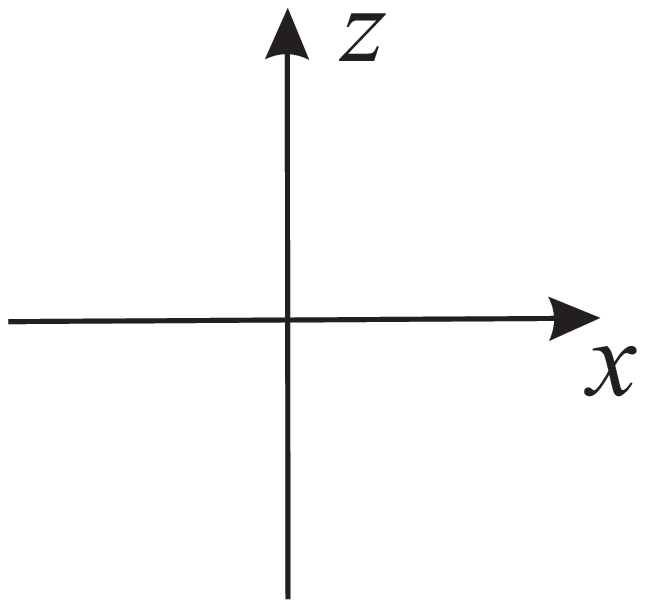}}
\caption{The phase portrait restricted to the region $x^2-z^2\leq1$ on the invariant plane $s=0$ inside the Poincar\'e disc.}
\label{Fig}
\end{minipage}
\end{figure}
\subsection{The invariant surface $x^2-z^2=1$}
\par Under the flow of system (\ref{eq1}), we first verify that $x^2-z^2=1$ is the invariant surface. If $l=l(x,z,s)=x^2-z^2-1$, then if the surface $x^2-z^2=1$ is invariant, then the polynomial $\mathcal{P}$ is required to make the equation
$$\frac{\partial l}{\partial x}\dot{x}+\frac{\partial l}{\partial z}\dot{z}+\frac{\partial l}{\partial s}\dot{s}=\mathcal{P}l,$$
true, which is exactly the case of $\mathcal{P}=2x(3x-\sqrt{6}s)$. 
\par On the surface $x^2-z^2=1$ system (1) writes
\begin{equation}
\begin{array}{rl}
\dfrac{dx}{dt}=3x\left(x^2-1\right),\ 
\dfrac{ds}{dt}=\dfrac{\sqrt{6}}{n}\,xs^2.
\end{array}
\end{equation}
Then except for $s$-axis is filled with equilibrium points, the above system also has two finite semi-hyperbolic equilibrium points $e_{1}=(1,0)$ and $e_{2}=(-1,0)$. It can be followed from Theorem 2.19 of \cite{Dumortier} that both $e_{1}$ and $e_{2}$ are saddle-nodes.
\par System (16) on the local chart $U_1$ writes 
\begin{equation}
\begin{array}{rl} 
\dfrac{du}{dt}=u\left[\dfrac{\sqrt{6}}{n}u+3\left(v^2-1\right)\right],\ 
\dfrac{dv}{dt}=3v\left(v^2-1\right).
\end{array}
\end{equation}
This system admits two infinite hyperbolic equilibrium points $e_{5}=(0,0)$ and $e_{6}=(\sqrt{6}n/2,0)$, where $e_{5}$ is a stable node and has eigenvalues $-3$ of multiplicity two, and the other point $e_{6}$ is an unstable saddle and has two eigenvalues $\pm3$. 
\par System (16) on the local chart $U_2$ can be written as
\begin{equation}
\begin{array}{rl}
\dfrac{du}{dt}=u\left[-\dfrac{\sqrt{6}}{n}u+3\left(u^2-v^2\right)\right],\ 
\dfrac{dv}{dt}=-\dfrac{\sqrt{6}}{n}\,uv.
\end{array}
\end{equation}
Rescaling the time $d\tau_4=udt$ we have
\begin{equation}
\begin{array}{rl}
\dfrac{du}{d\tau_4}=-\dfrac{\sqrt{6}}{n}u+3\left(u^2-v^2\right),\ 
\dfrac{dv}{d\tau_4}=-\dfrac{\sqrt{6}}{n}\,v.
\end{array}
\end{equation}
The origin $(0,0)$ of the above system is a hyperbolic stable node with eigenvalues $-\sqrt{6}/n$ and a multiplicity of 2. In this way, the origin $e_{7}=(0,0)$ of system (18) has a local phase portrait as shown in Figure 4.
\begin{figure}[htbp]
\begin{minipage}[t]{120mm}
\vspace {2mm}
\centering\includegraphics[width=6cm]{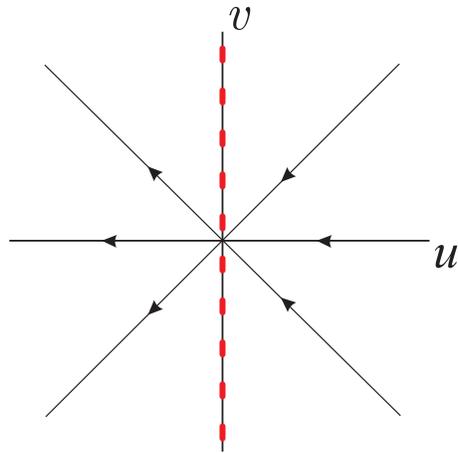}
\caption{The local phase portrait at the origin of $U_2$ for $x^2-z^2=1$.}
\label{Fig}
\end{minipage}
\end{figure}
\par In summary the global phase portraits of system (16) is integrated in Figure 5.
\begin{figure}[htbp]
\begin{minipage}[t]{120mm}
\vspace {2mm}
\centering\includegraphics[width=9.5cm]{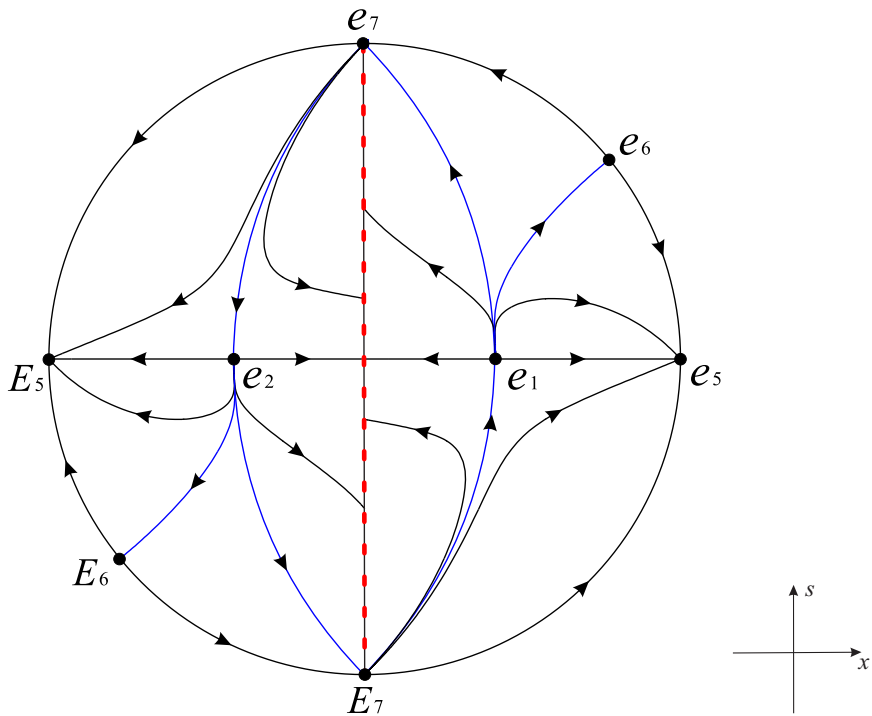}
  \subfigure{\includegraphics[width=2cm]{Fig1b5b_frame_x_s.eps}}
\caption{The phase portrait on the invariant surface $x^2-z^2=1$.}
\label{Fig}
\end{minipage}
\end{figure}
\subsection{The finite equilibrium points}
System (\ref{eq1}) allows three three-dimensional finite equilibrium points $p_0=(0,0,0)$, $p_1=(1,0,0)$ and $p_2=(-1,0,0)$, $p_0$ has eigenvalues $-3,0,0$, $p_1$ and $p_2$ have the same eigenvalues $6,3,0$. Here $p_1$ and $p_2$ are the intersection points of $x^2-z^2=1$, $s = 0$ and $z = 0$ that were just studied in the previous subsections 3.1-3.3, that is, $p_1$ and $p_2$ are the equilibrium points $e_1$ and $e_2$, respectively.
The origin $p_0$ is located in the middle of the intersection of $z = 0$ and $s = 0$, and it is the equilibrium point $e_0$ studied in the previous subsection 3.1.

\section{Phase portrait on the surface of Poincar\'e ball at infinity}
The three-dimensional Poincar\'e compactification (see Appendix or \cite{Cima} for more details) is used to study the dynamics of the system (\ref{eq1}) near infinity in this section. So we have $x=1/z_3,\ z=z_1/z_3,\ s=z_2/z_3$ on the local chart $U_1$, and then system (1) on the $U_1$ is reduced to
\begin{equation}
\begin{array}{rl} 
\dfrac{dz_1}{dt}&=z_1\left[3z_3^2 - \sqrt{6}z_2\left(-1 + z_1^2 + z_3^2\right)\right],\\
\dfrac{dz_2}{dt}&=z_2\left[3\left(z_3^2-1\right) + \sqrt{6}z_2\left(\dfrac{1+n}{n}  - z_1^2-z_3^2\right)\right],\\
\dfrac{dz_3}{dt}&=z_3\left[3\left(z_3^2-1\right) + \sqrt{6} z_2\left(1-z_1^2- z_3^2\right)\right].
\end{array}
\end{equation}
\par In different local charts, $z_3=0$ corresponds to the infinity of $\mathbb{R}^3$. The equilibrium points of the system (20) are listed in Table 1, where the equilibrium point $u_{31}$ represents the origin of the local chart $U_3$, and the other equilibrium points lie in the local chart $U_1$.
Additionally, for any constant $a$, $u_{a0}$ means that $s=0$ on local chart $U_1$ is filled with equilibrium points.
\begin{table}[!htb]
\newcommand{\tabincell}[2]{\begin{tabular}{@{}#1@{}}#2\end{tabular}}
\centering
\caption{\label{opt}The equilibrium points on the local charts of the surface of Poincar\'e ball at $\mathbb{R}^3$ infinity.}
\footnotesize
\rm
\centering
\begin{tabular}{@{}*{12}{l}}
\specialrule{0em}{2pt}{2pt}
 \toprule
\hspace{8mm}\textbf{Equilibrium points}&\textbf{Eigenvalues}\\
\specialrule{0em}{2pt}{2pt}
\toprule
\tabincell{l}{\hspace{8mm}$u_{11}=(0,0,0)$}&\tabincell{l}{$-3,-3,0$}\\
\specialrule{0em}{2pt}{2pt}
\hline
\specialrule{0em}{2pt}{2pt}
\tabincell{l}{\hspace{8mm}$u_{12}=\left(-1,\dfrac{\sqrt{6}}{2}n,0\right)$}&\tabincell{l}{$-3,3,-6n$}\\
\specialrule{0em}{2pt}{2pt}
\hline
\specialrule{0em}{2pt}{2pt}
\tabincell{l}{\hspace{8mm}$u_{13}=\left(1,\dfrac{\sqrt{6}}{2}n,0\right)$}&\tabincell{l}{$-3,3,-6n$}\\
\specialrule{0em}{2pt}{2pt}
\hline
\specialrule{0em}{2pt}{2pt}
\tabincell{l}{\hspace{8mm}$u_{14}=\left(0,\dfrac{\sqrt{6}n}{2(n+1)},0\right)$}&\tabincell{l}{$-\dfrac{3}{n+1},\dfrac{3n}{n+1},3$}\\
\specialrule{0em}{2pt}{2pt}
\hline
\specialrule{0em}{2pt}{2pt}
\tabincell{l}{\hspace{8mm}$u_{a0}=(a,0,0)$}&\tabincell{l}{$-3,-3,0$}\\
\specialrule{0em}{2pt}{2pt}
\hline
\specialrule{0em}{2pt}{2pt}
\tabincell{l}{\hspace{8mm}$u_{31}=(0,0,0)$}&\tabincell{l}{$0,0,0$}\\
\specialrule{0em}{2pt}{2pt}
 \toprule
\end{tabular}
\end{table}

For the case $z_3=0$ system (20) becomes
\begin{equation}
\begin{array}{rl} 
\dfrac{dz_1}{dt}&=\sqrt{6}z_1z_2\left(1- z_1^2 \right),\\
\dfrac{dz_2}{dt}&=z_2\left[-3 + \sqrt{6}z_2\left(\dfrac{1+n}{n} - z_1^2\right)\right].
\end{array}
\end{equation}
Rescaling the time $d\tau_5=z_2dt$, system (20) is reduced to
\begin{equation}
\begin{array}{rl} 
\dfrac{dz_1}{d\tau_5}&= \sqrt{6}z_1\left(1 - z_1^2 \right),\\
\dfrac{dz_2}{d\tau_5}&=-3 + \sqrt{6}z_2\left(\dfrac{1+n}{n}  - z_1^2\right).
\end{array}
\end{equation}
Then system (22) allows equilibrium points $e_{i,1}$, $e_{i,2}$ and $e_{i,3}$, the coordinates of which are $(\mp1,\sqrt{6}n/2)$ and $(0,\sqrt{6}n/(2n+2))$, respectively. The equilibrium points $e_{i,1}$ and $e_{i,2}$ are hyperbolic unstable saddles with eigenvalues $3$ and $-6n$, and the equilibrium point $e_{i,3}$ is a hyperbolic unstable node with eigenvalues  $3n/(1+n)$ and $3$. The phase portrait on local chart $U_1$ of the Poincar\'e sphere at infinity is shown in Figure 6.
\begin{figure}[htbp]
\begin{minipage}[t]{120mm}
\vspace {2mm}
\centering\includegraphics[width=9.5cm]{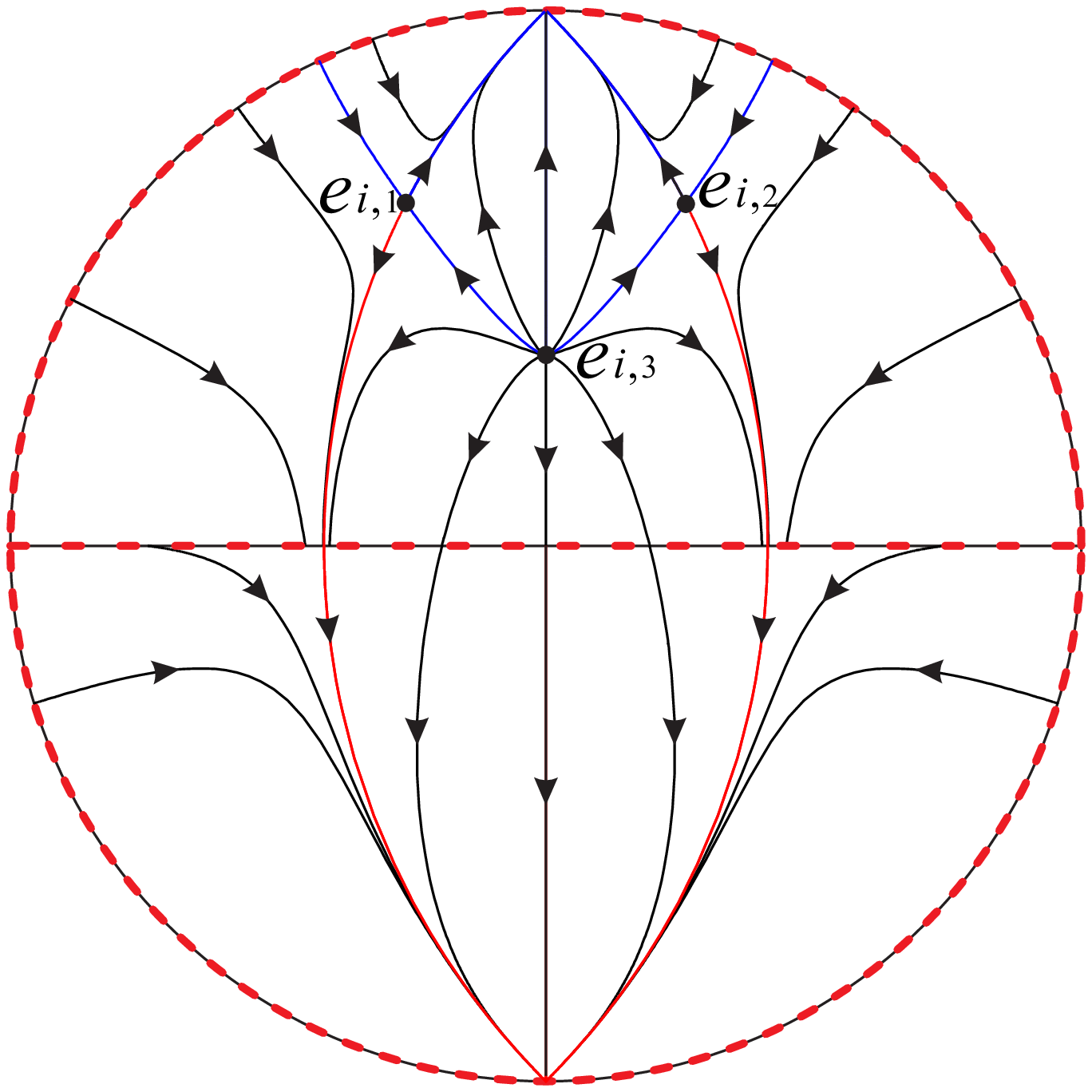}
  \subfigure{\includegraphics[width=2cm]{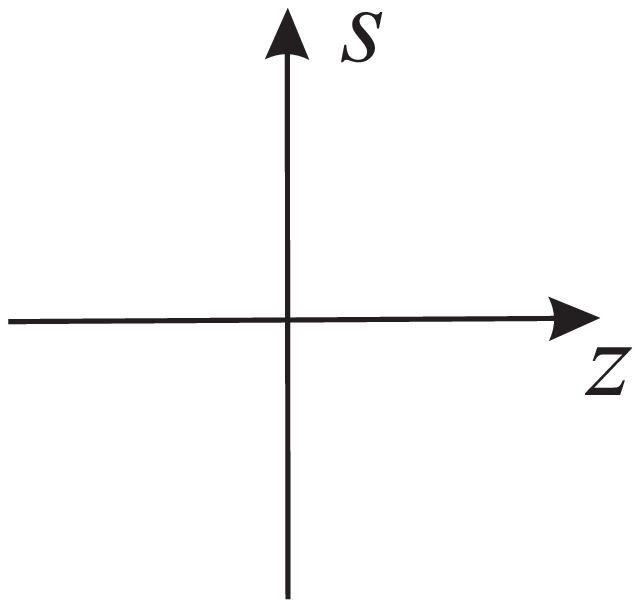}}
\caption{The phase portrait of system (\ref{eq1}) on the local chart $U_1$ at infinity.}
\label{Fig}
\end{minipage}
\end{figure}

On the local chart $U_2$ we have Poincar\'e compactification $x=z_1/z_3,\ z=1/z_3,\ s=z_2/z_3$, the system (1) becomes
\begin{equation}
\begin{array}{rl} 
\dfrac{dz_1}{dt}&=-z_1 z_3^2 + \sqrt{6} z_2 \left(1 - z_1^2 + z_3^2\right),\\
\dfrac{dz_2}{dt}&=z_1z_2 \left(- 3 z_1+ \dfrac{\sqrt{6}}{n} z_2 \right),\\
\dfrac{dz_3}{dt}&= - 3z_1^2 z_3.
\end{array}
\end{equation}

To study the phase portrait at infinity we take $z_3=0$, and changing the time $d\tau_6=z_2dt$ system (23) is equivalent to
\begin{equation}
\begin{array}{rl} 
\dfrac{dz_1}{d\tau_6}&= \sqrt{6} \left(1 - z_1^2 \right),\\
\dfrac{dz_2}{d\tau_6}&=z_1\left(- 3 z_1+ \dfrac{\sqrt{6}}{n} z_2 \right).
\end{array}
\end{equation}
Since $(0, 0)$ is the non-equilibrium point of system (24), there is no need to continue to investigate the equilibrium points at infinity in $U_2$. These have been discussed in the chart $U_1$.

On the local chart $U_3$ we have Poincar\'e compactification $x=z_1/z_3,\ z=z_2/z_3,\ s=1/z_3$, then system (1) writes
\begin{equation}
\begin{array}{rl} 
\dfrac{dz_1}{dt}&=z_1 \left[-\dfrac{\sqrt{6}(1+n)}{n} z_1 + 3 z_1^2 - 3 z_3^2\right] + \sqrt{6} \left(z_2^2 + z_3^2\right),\\
\dfrac{dz_2}{dt}&=z_1z_2 \left(-\dfrac{\sqrt{6}}{n} + 3 z_1\right),\\
\dfrac{dz_3}{dt}&=-\dfrac{\sqrt{6}}{n} z_1z_3.
\end{array}
\end{equation}

For the case $z_3=0$ it can be followed from system (25) that
\begin{equation}
\begin{array}{rl} 
\dfrac{dz_1}{dt}&=z_1 \left[-\dfrac{\sqrt{6}(1+n)}{n} z_1 + 3 z_1^2 \right] + \sqrt{6}z_2^2,\\
\dfrac{dz_2}{dt}&=z_1z_2 \left(-\dfrac{\sqrt{6}}{n} + 3 z_1\right).
\end{array}
\end{equation}
Note that the origin $(0,0)$ is a linearly zero equilibrium point of the above system. According to the Poincar\'e-Hopf Theorem, the topological index is zero. The vertical blow-up technique will be applied to investigate its local phase portrait. Then doing $w=z_2/z_1$ we have
\begin{equation}
\begin{array}{rl} 
\dfrac{dz_1}{dt}&=z_1^2 \left[\sqrt{6}\left(-\dfrac{1+n}{n}  + w^2\right) + 3z_1\right],\\
\dfrac{dw}{dt}&=\sqrt{6} z_1w \left(w^2-1\right).
\end{array}
\end{equation}
Rescaling the time $d\tau_7=z_1dt$ and eliminating the common factor $z_1$ of system (27), then we obtain
\begin{equation}
\begin{array}{rl} 
\dfrac{dz_1}{d\tau_7}&=z_1\left[\sqrt{6}\left(-\dfrac{1+n}{n}  + w^2\right) + 3z_1\right],\\
\dfrac{dw}{d\tau_7}&=\sqrt{6} w \left(w^2-1\right).
\end{array}
\end{equation}
Since there are three hyperbolic equilibrium points $e_{i,4}=(0,-1)$, $e_{i,5}=(0,1)$ and $e_{i,6}=(0,0)$ of system (28) on $z_1=0$, the previous two are stable nodes with eigenvalues of $-2\sqrt{6}$ and $-\sqrt{6}/n$, the last one is an unstable saddle point with eigenvalues of $-\sqrt{6}(n+1)/n$ and $\sqrt{6}$. We note that this is the same as the equilibrium points in the system (26) of reference \cite{Gao20192}, so the local phase portraits of systems (28), (27) and (26) are shown in Figures 7(a), 7(b) and 7(c), respectively. Thus Figure 8 shows the phase portrait at the origin of the local chart $U_3$. 
\begin{figure}[]
  \centering
  \begin{minipage}{130mm}
  \subfigure[]{\label{fig:subfig:a}
    \includegraphics[width=6cm]{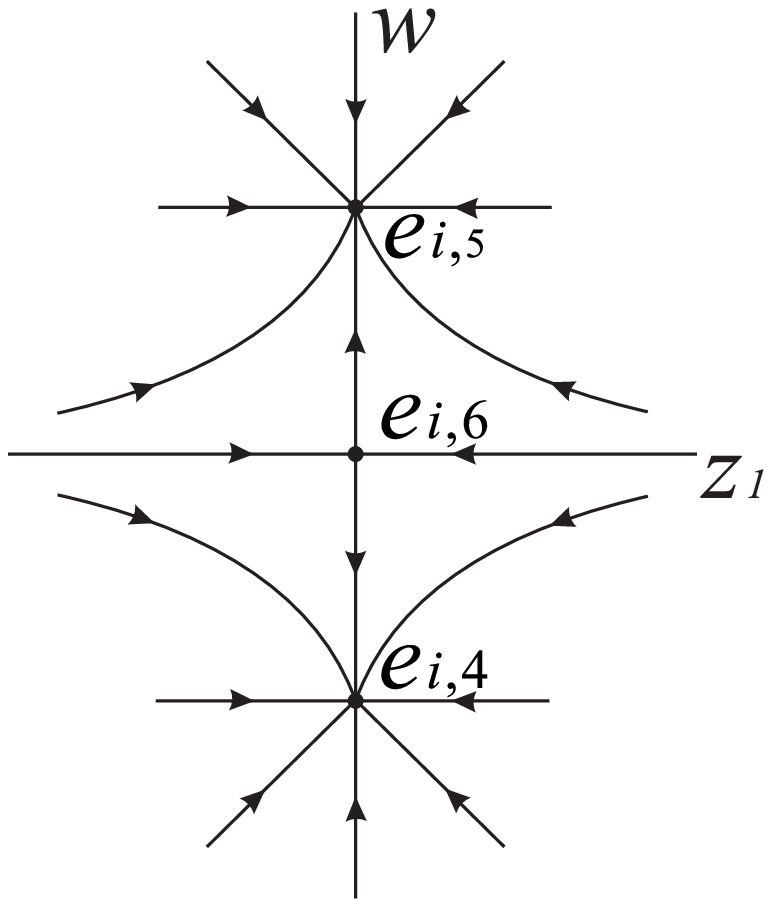}}
  \subfigure[]{\label{fig:subfig:b}
    \includegraphics[width=6cm]{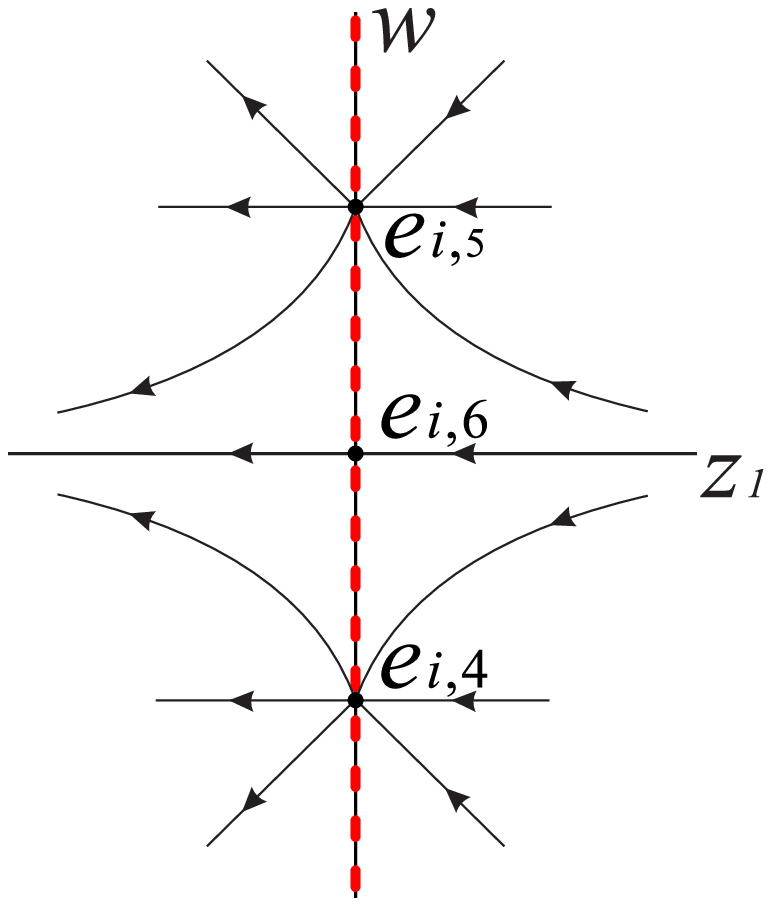}}\\
  \centering\subfigure[]{\label{fig:subfig:c}
    \includegraphics[width=6cm]{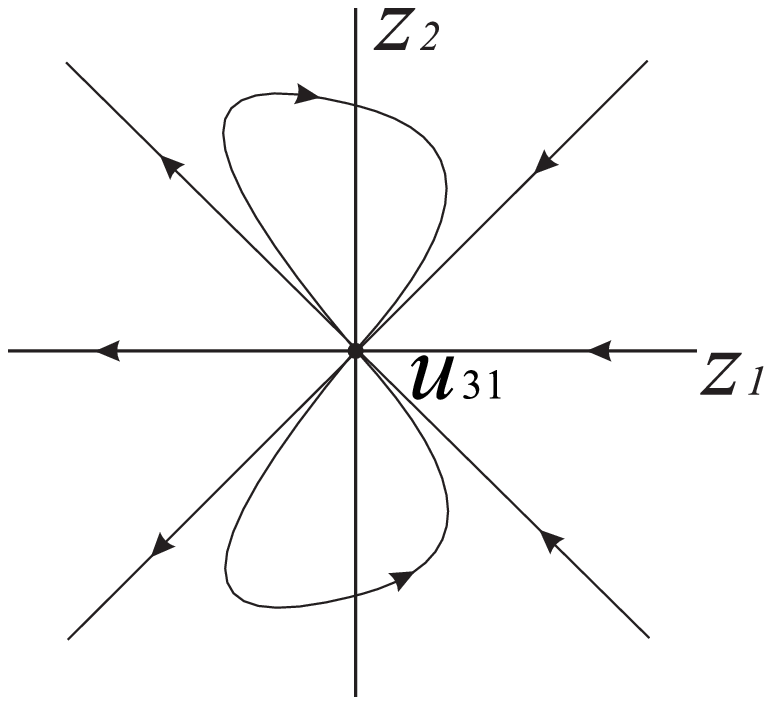}}
    \caption{The local phase portraits (a), (b) and (c) of the above equilibrium points correspond to systems (28), (27) and (26) respectively.}
  \label{fig:subfig}
  \end{minipage}
 \end{figure}
\begin{figure}[htbp]
\begin{minipage}[t]{120mm}
\vspace {2mm}
\centering\includegraphics[width=9.5cm]{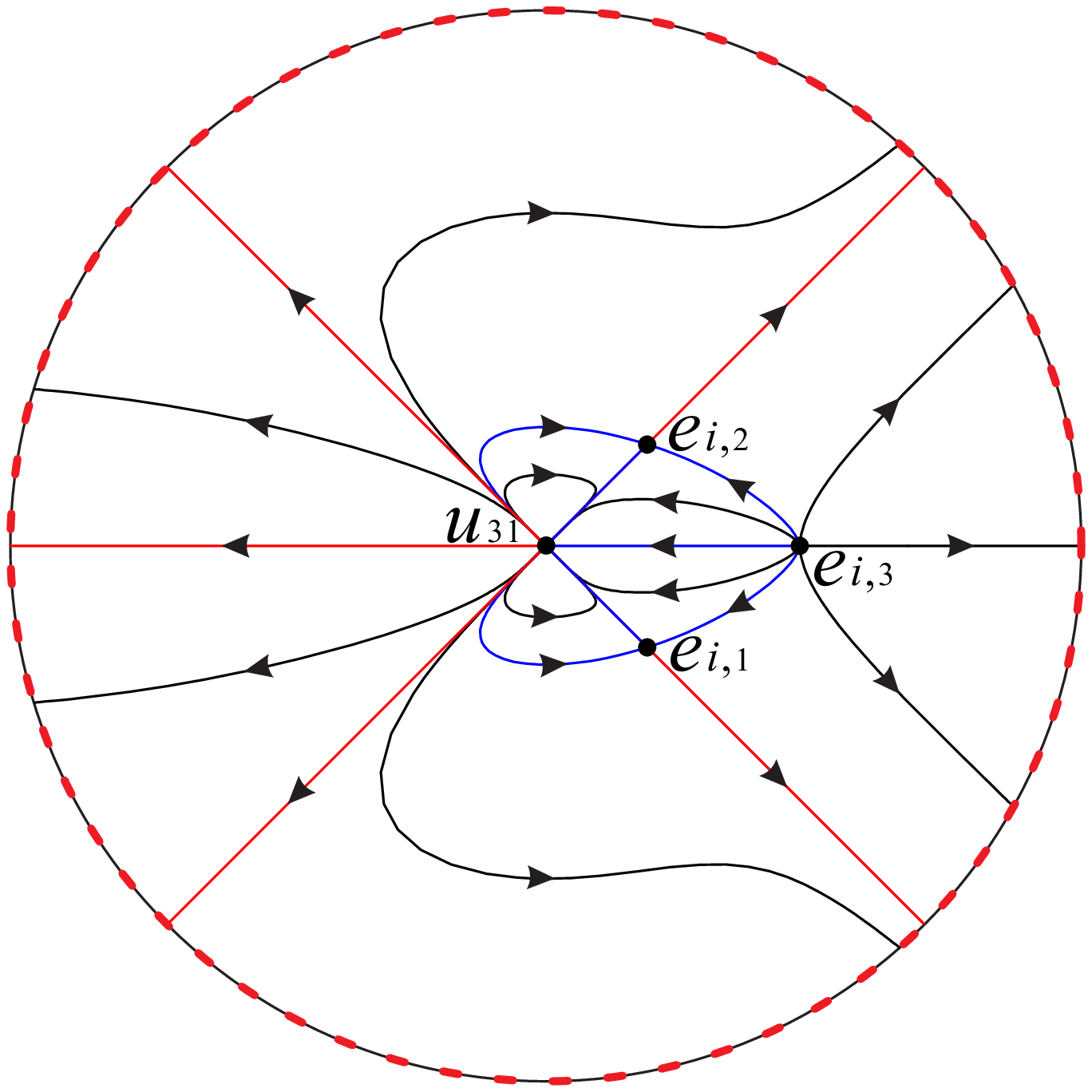}
  \subfigure{\includegraphics[width=2cm]{Fig3b8b_frame_x_z.eps}}
\caption{The phase portrait of system (\ref{eq1}) in the local chart $U_3$ at infinity.}
\label{Fig}
\end{minipage}
\end{figure} 
\par 
Combined with the previous discussion, Figure 9 shows the global phase portrait at infinity on the Poincar\'e sphere.
\begin{figure}[htbp]
\begin{minipage}[t]{120mm}
\vspace {2mm}
\centering\includegraphics[width=9.5cm]{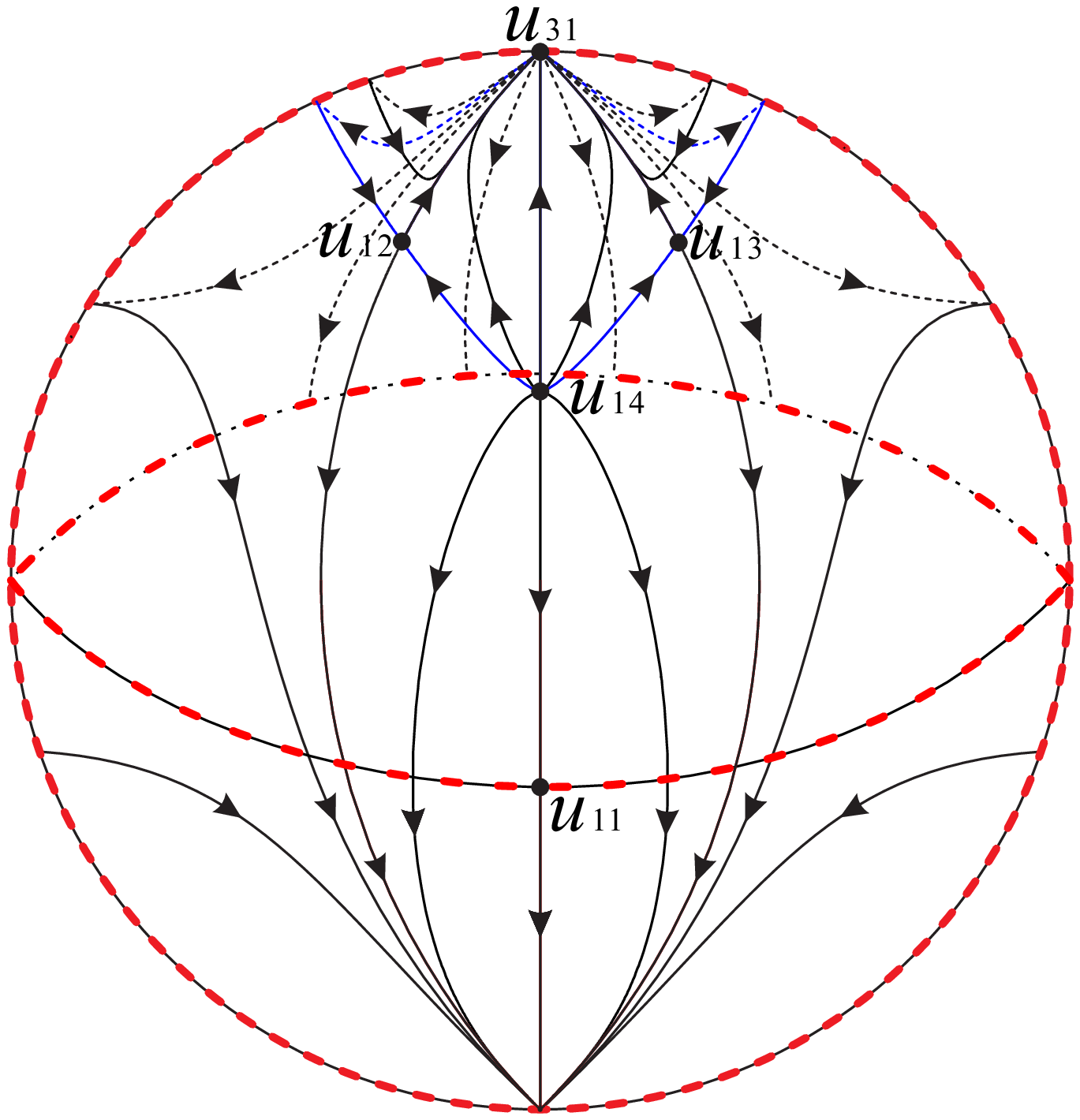}
  \subfigure{\includegraphics[width=2cm]{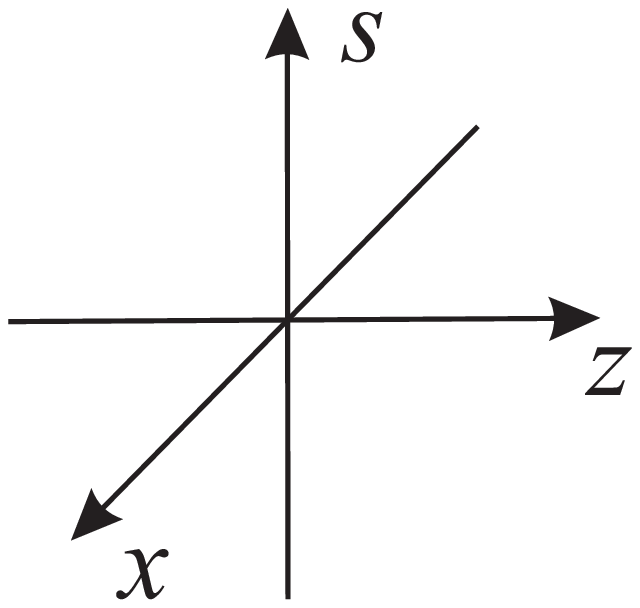}}
\caption{The global phase portrait of system (\ref{eq1}) on the Poincar\'e sphere at infinity.}
\label{Fig}
\end{minipage}
\end{figure} 

\section{Phase portraits within the Poincar\'e sphere conditioned to $x^2-z^2\leq1$}
\par Since system (\ref{eq1}) is invariant under the two symmetry about the origin and the $z$-axis, i.e., $(x,z,s)\mapsto(-x,-z,-s)$ and $(x,z,s)\mapsto(-x,z,-s)$. So we divide
the Poincar\'e ball restricted to the region $G$ into four regions as follows
\begin{equation*}
\begin{array}{rl} 
R_1:\ z\leq0,\ s\geq0.\ \ \
R_3:\ z\geq0,\ s\geq0.\\
R_2:\ z\leq0,\ s\leq0.\ \ \ 
R_4:\ z\geq0,\ s\leq0.
\end{array}
\end{equation*}
In view of the aforementioned symmetries, we only need to focus on the phase portrait of system (\ref{eq1}) in one region (such as $R_1$).
\par Combining the phase portrait in the invariant planes $s=0$, and $z=0$ with the phase portrait in the invariant surface $x^2-z^2=1$, and the phase portrait at infinity, the phase portrait on the boundary surface of $R_1$ is obtained in Figures 10-12.
\begin{figure}[htbp]
\begin{minipage}[t]{120mm}
\vspace {2mm}
\centering\includegraphics[width=6cm]{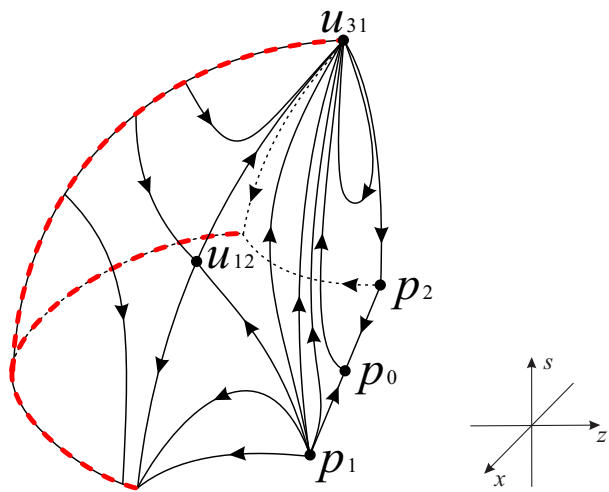}
  \subfigure{\includegraphics[width=2cm]{Fig9b10b11b12b14b15b_frame_x_z_s.eps}}
\caption{Phase portrait on the front boundary surface of $R_1$.}
\label{Fig}
\end{minipage}
\end{figure}
\begin{figure}[htbp]
\begin{minipage}[t]{120mm}
\vspace {2mm}
\centering\includegraphics[width=6cm]{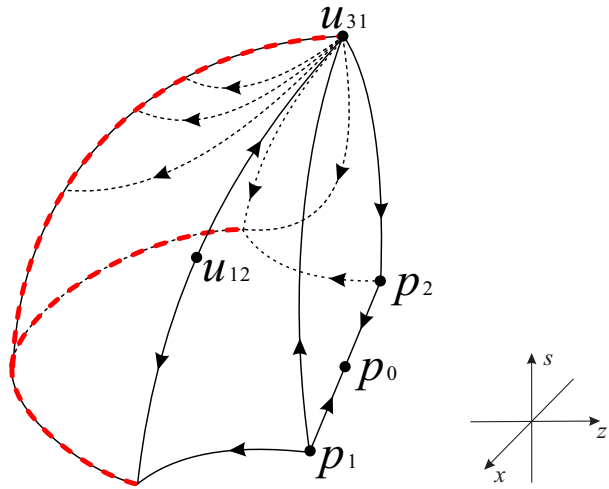}
  \subfigure{\includegraphics[width=2cm]{Fig9b10b11b12b14b15b_frame_x_z_s.eps}}
\caption{Phase portrait on the back boundary surface of $R_1$.}
\label{Fig}
\end{minipage}
\end{figure}
\begin{figure}[htbp]
\begin{minipage}[t]{120mm}
\vspace {2mm}
\centering\includegraphics[width=6cm]{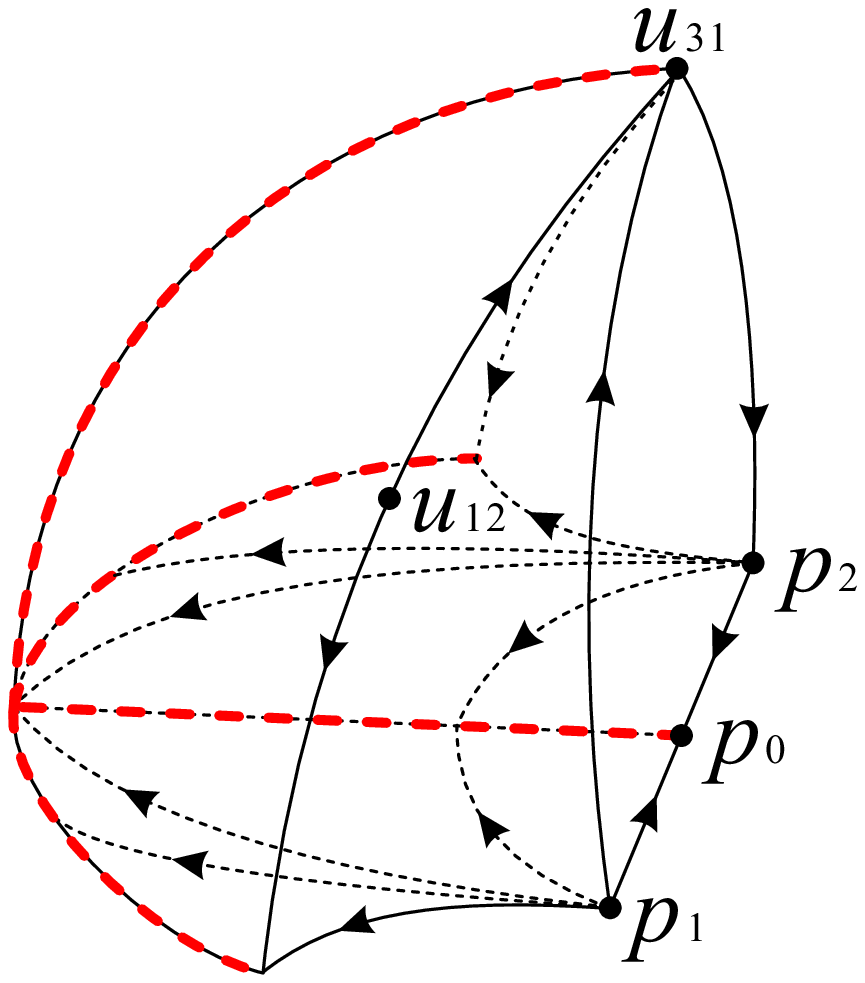}
  \subfigure{\includegraphics[width=2cm]{Fig9b10b11b12b14b15b_frame_x_z_s.eps}}
\caption{Phase portrait in the bottom plane of $R_1$.}
\label{Fig}
\end{minipage}
\end{figure}

\par Now we divide the boundary surface of the region $R_1$ into six parts (see Figure 13 for more details), thus the phase portrait of $R_1$ will be shown more clearly. It can be found from Figures 10 and 11 that the equilibrium point $u_{31}$ of the Poincar\'e ball is stable on the front boundary surfaces $F_1$ and $F_2$, and there is a stable parabolic sector and an elliptic sector segment $F_3$. However, on the back boundary surfaces $B_1$ and $B_2$, the north pole $u_{31}$ is unstable. 
\begin{figure}[htbp]
\begin{minipage}[t]{130mm}
\vspace {2mm}
\centering\includegraphics[width=8cm]{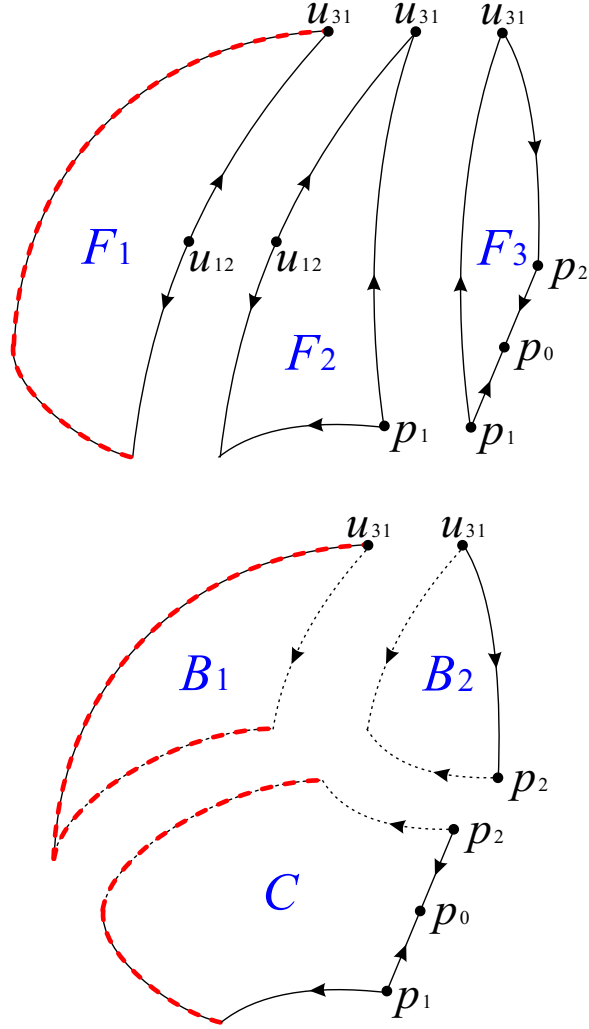}
\caption{The six boundaries of $R_1$.}
\label{Fig}
\end{minipage}
\end{figure}

\section{Dynamics inside the region $R_1$}

\par System (\ref{eq1}) has three finite equilibrium points $p_0$, $p_1$ and $p_2$. The dynamical behavior of system (\ref{eq1}) in the Interior of the region $R_1$ depends on the comprehensive performance of the flow in the surface and planes 
$$h(x,z,s)=0,\ x=0,\ z=0,\ s=0,$$
where
\begin{equation*}
\begin{array}{rl} 
h(x,z,s)&=\sqrt{6}s\left(z^2-x^2+1\right)+3x\left(x^2-1\right).
\end{array}
\end{equation*}
\par The above planes and surface cut the region $R_1$ into four different subregions $S_i,\ i=(1,\dots,4)$, see Figures 14 and 15 for more details. It is noted that $h<0$ in the subregions $S_1$ and $S_4$, and $h>0$ in the subregions $S_2$ and $S_3$. To avoid confusion, please note that the solid and dotted lines in Figure 15 is consistent with those in Figure 14.
\begin{figure}[htbp]
\begin{minipage}[t]{120mm}
\vspace {2mm}
\centering\includegraphics[width=7cm]{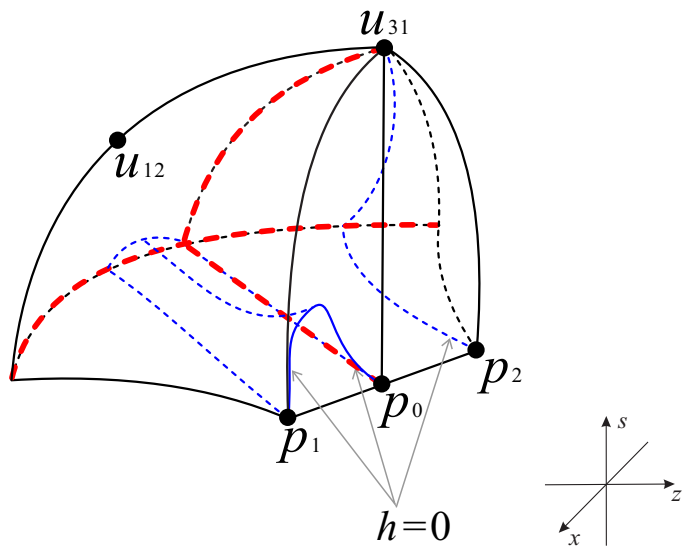}
  \subfigure{\includegraphics[width=2cm]{Fig9b10b11b12b14b15b_frame_x_z_s.eps}}
\caption{There are four subregions inside the region $R_1$ of the Poincar\'e ball.}
\label{Fig}
\end{minipage}
\end{figure}
\begin{figure}[]
  \centering
  \begin{minipage}{130mm}
  \subfigure[]{\label{fig:subfig:a}
    \centering\includegraphics[width=4cm]{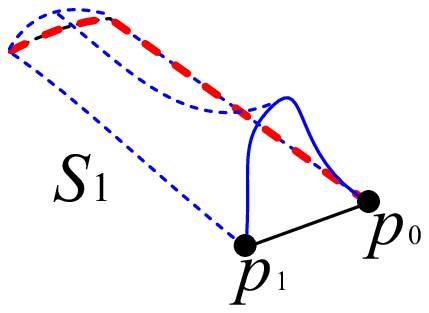}}
  \subfigure[]{\label{fig:subfig:b}
    \centering\includegraphics[width=5.5cm]{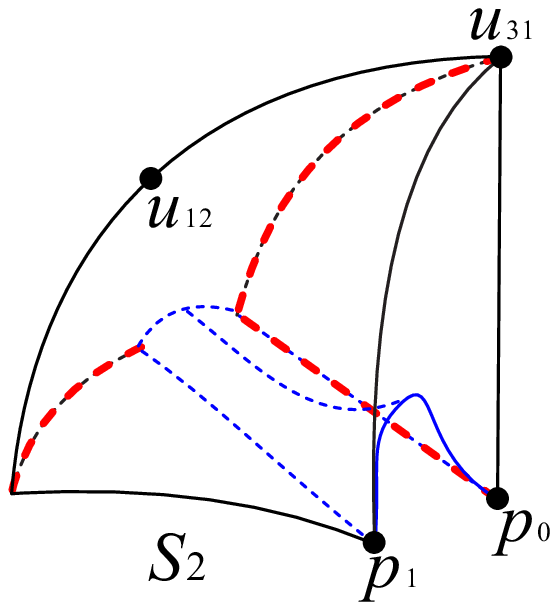}}\\
  \subfigure[]{\label{fig:subfig:c}
    \centering\includegraphics[width=4.5cm]{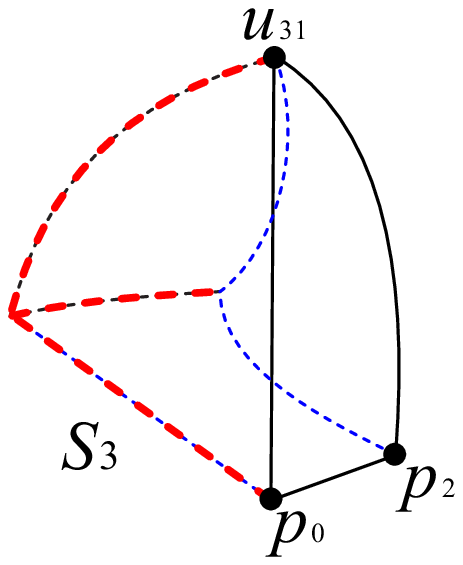}}
  \hspace{1cm}
  \subfigure[]{\label{fig:subfig:d}
    \centering\includegraphics[width=2.5cm]{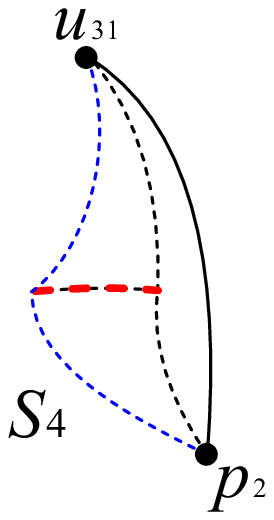}}
    \centering\subfigure{\includegraphics[width=2cm]{Fig9b10b11b12b14b15b_frame_x_z_s.eps}}
    \caption{The four subregions $S_i$ of $R_1\ (i=1\cdots,4)$.}
  \label{fig:subfig}
  \end{minipage}
 \end{figure}

\par As it is shown in $S_1$ (see Figure 15(a)) the upper surface is included in the blue surface $h = 0$, and the bottom plane is included in the invariant plane $s = 0$. According to Table 2, the orbits monotonically decrease in the $x$ and $z$ directions. In the $s$ direction, they monotonically increase, so the orbits in the subregion $S_1$ can only start at the finite equilibrium point $p_1$ and pass through the upper surface of $S_1$ into the subregion $S_2$. However in subregion $S_2$ (see Figure 15(b), i.e. the remaining part after $S_1$ is extracted from the region $R_1$ when $x>0$), the orbits monotonically increase in the $x$ and $s$ directions but monotonically decrease in the $z$-direction, so we find that the orbits in $S_2$ can only start at $p_1$ or the equilibrium points on the negative $z$-axis, and they eventually towards the infinity equilibrium point $u_{12}$ fixed on the surface of $S_2$. Therefore the dynamic behavior of the orbits on these two subregions can be represented in the following form
$$\begin{tikzpicture}[->, thick] 
    \node (u12) at (0, 0) [] {$u_{12}$};
        \node (S2) at (-1.5, 0) [] {$S_2$};
                \node (S1) at (-3, 0) [] {$S_1$};
                        \node (p1) at (-2.25, 1.25) [] {$p_1$};
         \node (z) at (2.7, 0) [] {negative $z$-axis.};   
\draw (p1) -- (S1) ;
\draw (p1) -- (S2) ;
\draw (S1) -- (S2) ;
\draw (S2) -- (u12) ;
\draw (z) -- (u12) ;
\end{tikzpicture}$$
\begin{table}[!htb]
\newcommand{\tabincell}[2]{\begin{tabular}{@{}#1@{}}#2\end{tabular}}
\centering
\caption{\label{opt}The dynamics of the four subregions.}
\footnotesize
\rm
\centering
\begin{tabular}{@{}*{12}{l}}
\specialrule{0em}{2pt}{2pt}
 \toprule
\hspace{2mm}\textbf{Subregions}&\textbf{Associated Region}&\textbf{Monotonicity}\\
\specialrule{0em}{2pt}{2pt}
\toprule
\tabincell{l}{\hspace{8mm}$S_1$}&\tabincell{l}{$h<0,\ x>0,\ z<0,\ s>0$}&\tabincell{l}{$\dot{x}<0,\ \dot{z}<0,\ \dot{s}>0$}\\
\specialrule{0em}{2pt}{2pt}
\hline
\specialrule{0em}{2pt}{2pt}
\tabincell{l}{\hspace{8mm}$S_2$}&\tabincell{l}{$h>0,\ x>0,\ z<0,\ s>0$}&\tabincell{l}{$\dot{x}>0,\ \dot{z}<0,\ \dot{s}>0$}\\
\specialrule{0em}{2pt}{2pt}
\hline
\specialrule{0em}{2pt}{2pt}
\tabincell{l}{\hspace{8mm}$S_3$}&\tabincell{l}{$h>0,\ x<0,\ z<0,\ s>0$}&\tabincell{l}{$\dot{x}>0,\ \dot{z}<0,\ \dot{s}<0$}\\
\specialrule{0em}{2pt}{2pt}
\hline
\specialrule{0em}{2pt}{2pt}
\tabincell{l}{\hspace{8mm}$S_4$}&\tabincell{l}{$h<0,\ x<0,\ z<0,\ s>0$}&\tabincell{l}{$\dot{x}<0,\ \dot{z}<0,\ \dot{s}<0$}\\
\specialrule{0em}{2pt}{2pt}
 \toprule
\end{tabular}
\end{table}  
\par In subregion $S_3$ (see Figure 15(c)) the left vertical plane facing us is contained in the plane $x = 0$, and the right vertical plane is contained in the invariant plane $z = 0$. In the opposite surfaces the left one is included in the surface of the Poincar\'e sphere, and the right one is included in the surface $h=0$. Since the orbits in $S_3$ monotonically decrease in the $z$ and $s$ directions, and they increase monotonically in the $x$ direction, it follows that the orbits will eventually go to the equilibrium point which lie in the equator or go to the negative $z$-axis in that region, and the orbits in this region may only come from the adjacent subregion $S_4$ (see Figure 15(d)). The left and right surfaces of subregion $S_4$ belong to the surfaces $h=0$ and $x^2-z^2=1\ (x<0)$, respectively. Since all the orbits are monotonically decreasing in $S_4$, the orbits in this region arrive from the infinite equilibrium point $u_{31}$, i.e. north pole of the Poincar\'e ball, and tend to the infinite equilibrium points on the equator of this region at a future date, or enter into the subregion $S_3$. This dynamical behavior can be represented as follows
$$\begin{tikzpicture}[->, thick] 
    \node (S4) at (0, 0) [] {$S_4$};
        \node (u31) at (-1.5, 0) [] {$u_{31}$};
        \node (eS4) at (0, 1.4) [] {equator in $S_4$};
                \node (S3) at (1.4, 0) [] {$S_3$};
                        \node (eS3) at (1.5, -1.4) [] {equator in $S_3$};
                                \node (z) at (4, 0) [] {negative $z$-axis.};
\draw (u31) -- (S4) ;
\draw (S4) -- (eS4) ;
\draw (S4) -- (S3) ;
\draw (S3) -- (eS3) ;
\draw (S3) -- (z) ;
\end{tikzpicture}$$
 
Therefore the orbits' dynamic behavior inside the four subregions of $R_1$ investigated above can be condensed to
$$\begin{tikzpicture}[->, thick] 
 \node (u31) at (0, 0) [] {$u_{31}$};
  \node (z) at (2.6, 0) [] {negative $z$-axis};
  \node (u12) at (5.25, 0) [] {$u_{12}.$};
  \node (p1) at (5.25, -1.3) [] {$p_1$};   
  \node (eS4) at (0, 1.3) [] {equator in $S_4$};
  \node (eS3) at (0, -1.3) [] {equator in $S_3$};                        
\draw (u31) -- (eS4) ;
\draw (u31) -- (eS3) ;
\draw (u31) -- (z) ;
\draw (z) -- (u12) ;
\draw (p1) -- (u12) ;
\end{tikzpicture}$$

This flow chart states clearly that the orbits of system (\ref{eq1}) included in $R_1$ have $\alpha$-limit at $p_1$ and the north pole $u_{31}$ (also referred to as past attractors or negative attractors) of the Poincar\'e sphere. Additionally the orbits have $\omega$-limit either at $u_{12}$ (also called future attractor or positive attractor), or other infinite equilibrium points on the equator of subregions $S_3$ and $S_4$, which are located at the intersection of the Poincar\'e ball and at the infinity of the invariant planes $s=0$ in $\mathbb{R}^3$ when $x<0$, see Figure 14 or Figures 15(c) and 15(d).

Therefore all the global dynamical behavior of the system (\ref{eq1}) is represented qualitatively and completely.

 \section{Conclusions}
 
\par In the present paper we have fully described the global phase portrait of Ho\v{r}ava-Lifshitz cosmology in the presence of non-zero cosmological constant and zero curvature in the region of physical interest $G$. By taking the fact that the cosmological equations remains invariant under the two symmetries mentioned in section 5, the global phase portrait of the cosmological model in $G$ is provided completely. 

From the perspective of cosmology, combined with the previous analysis of the phase portraits of the system, we know that the unstable finite equilibrium points $p_1$ and $p_2$ are dominated by dark matter, and the finite equilibrium $p_0$ located on the equilibrium point line may also be characterized by dark matter if the initial conditions are not in the invariant plane $z = 0$. Besides the initial conditions in the invariant planes $s = 0$ and $z=0$ as well as on the backside of the invariant surface $x^2-z^2=1$, the phase portrait displays that the eventual evolution of the orbits of the cosmological model in $G$ tends to the infinite equilibrium point $u_{12}$, which can be the late-time state of the universe and to other infinite equilibrium points placed at the equator of the Poincar\'e ball. For the Ho\v{r}ava-Lifshitz gravity in a FLRW space-time with $k=0$ and $\Lambda\neq0$, equations (\ref{eq5}) implies that the Hubble parameter $H$ tends to nil in forward time in this cosmological model.

\section*{Appendix: The Poincar\'{e} compactification in $\mathbb{R}^3$}For a polynomial vector field \textbf{$X$}$=(P_1,P_2,P_3)$, and the degree $n=\max{\{\deg(P_i): i=1,2,3\}}$, its differential system is
\begin{equation*}
\begin{array}{rl} 
\dfrac{dx}{dt}=P_1(x,y,z),\ \dfrac{dy}{dt}=P_2(x,y,z),\ \dfrac{dz}{dt}=P_3(x,y,z).
\end{array}
\label{eq}
\end{equation*}
\indent Defining the unit sphere in $\mathbb{R}^4$ by $\mathbb{S}^3={\{y=(y_1,y_2,y_3,y_4)\in\mathbb{R}^4:}$ ${\|y\|=1\}}$, we denote the northern hemisphere by $\mathbb{S}_+={\{y\in\mathbb{S}^3:}$ ${y_4>0\}}$, the southern hemisphere by $\mathbb{S}_-={\{y\in\mathbb{S}^3:}$ ${y_4<0\}}$, the equator of $\mathbb{S}^3$ by the sphere by $\mathbb{S}^2={\{y\in\mathbb{S}^3: y_4=0\}}$, the tangent space at the point $y$ of $\mathbb{S}^3$ by $T_y\mathbb{S}^3$. Hence the tangent hyperplane $T_{(0,0,0,1)}\mathbb{S}^3=$ ${\{(x_1,x_2,x_3,1)}$ ${\in\mathbb{R}^4: (x_1,x_2,x_3)\in\mathbb{R}^3\}}$ is identified with $\mathbb{R}^3$. Moreover let
\begin{equation*}
\begin{array}{rl}
f_+: \mathbb{R}^3=T_{(0,0,0,1)}\mathbb{S}^3\to\mathbb{S}+,\ \ \ 
f_+(x)=\dfrac{1}{\Delta x}(x_1,x_2,x_3,1),
\end{array}
\label{eq}
\end{equation*}
and
\begin{equation*}
\begin{array}{rl}
f_-: \mathbb{R}^3=T_{(0,0,0,1)}\mathbb{S}^3\to\mathbb{S}-,\ \ \ 
f_-(x)=-\dfrac{1}{\Delta x}(x_1,x_2,x_3,1),
\end{array}
\label{eq}
\end{equation*}
be the two central projections, where $\Delta x=\left(\sum^3_{i=1}x^2_i+1\right)^{1/2}$. Then $f_+$ and $f_-$ are identified by $\mathbb{R}^3$ with the two hemispheres of $\mathbb{S}^3$. These two central projections clarify two transcripts of $X$, i.e. $Df_+\circ X$ in $\mathbb{S}_+$, and $Df_-\circ X$ in $\mathbb{S}_-$. Let $\widetilde{X}$ be the vector field on $\mathbb{S}^3\backslash\mathbb{S}^2 = \mathbb{S}_+\cup\mathbb{S}_-$.\\
\indent Now we analytically continue the vector field $\widetilde{X}(y)$ to the entire sphere $\mathbb{S}^3$ by $p(X)(y)=y^{n-1}_4\widetilde{X}(y)$. The continued vector field $p(X)$ is named \textit{Poincar\'{e} compactification} of $X$. We note that the infinity of $\mathbb{R}^3$ represented by $\mathbb{S}^2$ is invariant for the vector field $p(X)$. The compactification for polynomial vector fields in $\mathbb{R}^2$ was introduced by Poincar\'{e}, and one can find its extension to $\mathbb{R}^m$ from \cite{Cima}. Next we shall study the orthogonal projection of the closed $\mathbb{S}_+$ to $y_4 = 0$, which is a closed ball $B$ (designated Poincar\'{e} ball) with radius 1, its inner area is diffeomorphic to $\mathbb{R}^3$, and its boundary surface $\mathbb{S}^2$ is identified with $\mathbb{R}^3$ at infinity. \\
\indent Since $\mathbb{S}^3$ is a differentiable manifold, we must consider eight local charts $(U_i, F_i)$ and $(V_i, G_i)$ in order to study the dynamics of $p(X)$, where 
\begin{equation*}
\begin{array}{rl}
U_i = {\{y\in\mathbb{S}^3: y_i> 0\}},\ \ \ 
V_i = {\{y\in\mathbb{S}^3: y_i <0\}},
\end{array}
\label{eq}
\end{equation*}
and the diffeomorphisms 
\begin{equation*}
\begin{array}{rl}
F_i : U_i\to\mathbb{R}^3,\ \ \ G_i : V_i\to\mathbb{R}^3\ \ \ \text{for}\ i = 1, 2, 3, 4,
\end{array}
\label{eq}
\end{equation*}
are the central projections' inverses from the origin to the center of the tangent planes at the points $(\pm1, 0, 0, 0)$, $(0, \pm1, 0, 0)$, $(0, 0, \pm1, 0)$ and $(0, 0, 0, \pm 1)$, respectively. Then the expression of $p(X)$ in the local chart $U_1$ is
\begin{equation*}
\begin{array}{rl}
\dfrac{z^n_3}{(\Delta z)^{n-1}}(-z_1P_1+P_2,-z_2P_1+P_3,-z_3P_1),
\end{array}
\label{eq}
\end{equation*}
where $P_i=P_i(1/z_3,z_1/z_3,z_2/z_3)$. In the local chart $U_2$ we have
\begin{equation*}
\begin{array}{rl}
\dfrac{z^n_3}{(\Delta z)^{n-1}}(-z_1P_2+P_1,-z_2P_2+P_3,-z_3P_2),
\end{array}
\label{eq}
\end{equation*}
where $P_i=P_i(z_1/z_3,1/z_3,z_2/z_3)$. In the local chart $U_3$ we obtain
\begin{equation*}
\begin{array}{rl}
\dfrac{z^n_3}{(\Delta z)^{n-1}}(-z_1P_3+P_1,-z_2P_3+P_2,-z_3P_3),
\end{array}
\label{eq}
\end{equation*}
where $P_i=P_i(z_1/z_3,z_2/z_3,1/z_3)$. In the local chart $U_4$ we get
\begin{equation*}
\begin{array}{rl}
z_3^{n-1}(P_1,P_2,P_3),
\end{array}
\label{eq}
\end{equation*}
where $P_i=P_i(z_1,z_2,z_3)$. Furthermore the demonstration of $p(X)$ in the local chart $V_i$ is the same as in the $U_i$ multiplied by $(-1)^{n-1}$. \\
\indent In addition the above factor $1/(\Delta z)^{n-1}$ is omitted by doing a time rescaling when we use the expressions of the compactified vector field $p(X)$ in the local charts.

 \section*{Acknowledgments}
 \par The first author was supported by the National Natural Science Foundation of China (NSFC) through grants 12172322 and 11672259, the China Scholarship Council through grant 201908320086.
 \par The second author was supported by the Ministerio de Econom$\acute{\i}$a, Industria y Competitividad, Agencia Estatal de Investigaci\'on grants MTM2016-77278-P (FEDER) and MDM-2014-0445, the Ag$\grave{\text{e}}$ncia de Gesti\'o d'Ajuts Universitaris i de Recerca grant 2017SGR1617, and the H2020 European Research Council grant MSCA-RISE-2017-777911.
 
\noindent\textbf{Conflicts of Interest:} The authors declare no conflict of interest.

\end{document}